\documentclass[english,aps,prb,twocolumn]{revtex4-1}
\usepackage[english]{babel}
\usepackage{natbib}
\usepackage{amsmath}
\usepackage{amssymb}
\usepackage{graphicx}
\usepackage{indentfirst}

\makeatletter

\renewcommand{\emph}[1]{\textit{#1}}

\begin{document}

\title{Coulomb and exchange interaction effects on the exact two-electron dynamics in the Hong-Ou-Mandel interferometer based on Hall edge states}
\author{L. Bellentani$^{1}$, P. Bordone$^{1,2}$, X. Oriols$^{3}$ and A. Bertoni$^{2}$}
\affiliation{$^1$Dipartimento di Scienze Fisiche, Informatiche e Matematiche, Universit{\`a} degli Studi 
di Modena e Reggio Emilia, Via Campi 213/A, I-41125 Modena, Italy}
\affiliation{$^2$S3, Istituto Nanoscienze-CNR, Via Campi 213/A, 41125 Modena, Italy}
\affiliation{$^3$Departament d'Enginyeria Electr\`{o}nica, Universitat Aut\`{o}noma de Barcelona, 08193-Bellaterra (Barcelona), Spain}

\begin{abstract}
The electronic Hong-Ou-Mandel interferometer in the integer quantum Hall regime is an ideal system to probe the building up of quantum correlations between charge carriers and it has been proposed as a viable platform for quantum computing gates.
Using a parallel implementation of the split-step Fourier method, we simulated the antibunching of two interacting fermionic wave packets impinging on a quantum point contact. Numerical results of the exact approach are compared with a simplified theoretical model based on one-dimensional scattering formalism. We show that, for strongly-localized wave packets in a full-scale geometry, the Coulomb repulsion dominates over the exchange energy, this effect being strongly dependent on the energy broadening of the particles. We define analytically the spatial entanglement between the two regions
of the quantum point contact, and obtain quantitatively its entanglement-generation capabilities.
\end{abstract}


\maketitle

\section{Introduction}
In the electronic counterpart of the Hong-Ou-Mandel (HOM) experiment two indistinguishable electrons impinge on the opposite sides of an half-reflecting quantum point contact (QPC), acting as the beam splitter. Contrary to photons  \citep{Hong1987_PRL}, the antisymmetry of the two-electron wave function entails in ideal conditions a zero bunching probability  \citep{Buttiker1992_PRL, Buttiker1992_PRB}. However, in experiments  \citep{Bocquillon2013_S, Oliver1999_S} the energy-broadening of electrons induces a non zero Pauli dip. This has been related to the interplay between the energy selectivity of the QPC and the energetic dispersion of the wave packets  \citep{Marian2015_JPCM}, or to the decoherence induced by charge fractionalization  \citep{Marguerite2016_PRB, Freulon2015_NC}.

A reliable and robust HOM interferometer is essential in the flying qubit implementation of quantum computing gates based on Hall edge-state interferometers \citep{Emary2019_PRB}. Indeed, it requires a system that is invariant to small perturbations and robust against typical scattering mechanisms in semiconductor devices, such as with phonons, impurities and background electrons. Experiments  \citep{Rolleau2008_PRL, Kataoka2016_PRL} show that coherent transport of electrons can be achieved in the Integer Quantum Hall (IQH) regime  \citep{Sarma1997}, which is generated by an intense perpendicular magnetic field applied to a confined two-dimensional electron gas (2DEG). This produces \textit{edge states}, chiral 1D channels following the borders of the device, where an electron can be injected and propagates without being backscattered  \citep{Venturelli2011_PRB, Palacios1992_PRB}. Edge-channel based nanodevices have been implemented to test electron self-interference in the Mach-Zehnder interferometer \cite{Bellentani2018_PRB, Beggi2015_JOPCM, Levkivskyi2008_PRB, Bird1994_PRB, Karmakar2015_PRB, Neder2006_PRL, Rossello2015_PRB}, to observe violation of Bell's inequality and two-qubit correlations in the Hanbury-Brown Twiss interferometer  \citep{Samuelsson2004_PRL, Neder2007_N}, and to devise quantum logic gates as quantum erasers or which path detectors  \citep{Weisz2014_S, Neder2007_PRL, Radu2019_PHYSE}. Moreover, a protocol for the quantum tomography \cite{Jullien2014_Nature} of a fermionic particle has been proposed in a Hanbury-Brown Twiss interferometer \cite{Grenier2011_NJP}. 
Although fractional quantum Hall effect \cite{Laughlin1983_PRL} offers many additional opportunities to exploit coherent quasiparticle interference \cite{Wahl2014_PRL,Rech2017_PRL,Ronetti2018_PRB}, we limit our present study to the integer regime.
 
Regarding the electronic HOM effect, recent experiments highlight the presence of charge fractionalization \cite{Laughlin1983_PRL,Saminadayar1997_PRL,Kapfer2019_Science} in the propagation of single excitations in edge channels, so that the coherence of the travelling qubit is not preserved  \citep{Roussel_PSS, Marguerite_PSS}. This effect originates at bulk filling factor 2 due to inter-channel interactions that destroy the coherence of the injected Landau quasi-particles  \citep{Huyn2012_PRL, Ferraro2014_PRL, Helzel2015_PRB}. As recently proposed, strategies can be implemented to quench this source of decoherence, e.g. the introduction of top gates to loop the second channel \cite{Cabart2018_PRB} or to increase the distance between the two copropagating states  \citep{Bellentani2018_PRB}. 
Contrary to the previous scenario, in the present work, we analyze and simulate a device operating at bulk filling factor 1, i.e. only the first Landau level is energetically available  \citep{Beggi2015_JOPCS}. Rather than considering the injection of a steady-state current in the Hall interferometer \cite{Ji2003_N, Neder2007_N, Henny1999_S, Oliver1999_S}, we simulate our flying-qubit as encoded in a Gaussian wave packet of edge states \cite{Bellentani2018_PRB, Beggi2015_JOPCM}, with an energy well above the Fermi sea, as recently proposed using quantum dot pumps with time-dependent confining barriers  \citep{Ryu2016_PRL}. 
The choice of a Gaussian state minimizes the computational burden, since its spatial spreading during the propagation is lower with respect to other kinds of excitations, as Levitons \cite{Ivanov1997_PRB,Feve2007_Science,Dubois2013_Nature,Ferraro2018_EuPJST}. However, we expect that our results do not depend on the shape of the wave packets, at least at a qualitative level, as long as their components transmitted and reflected by the quantum point contact have a similar distribution \cite{Beggi2015_JOPCM}.

In our approach, the time-dependent wave function is evolved by means of a parallel implementation of the split-step Fourier method  \citep{Bellentani2018_PRB, Beggi2015_JOPCM} in a two-dimensional potential landscape, that reproduces the field generated by top gates in the typical GaAs/AlGaAs heterostructure. The energy broadening of the single-particle is directly included in the propagating state and time is an intrinsic variable of our simulations, so that we can access the dynamical properties of the two-particle system. It should be noticed that traditional approaches in the literature bypass the huge computational load for such simulations by using scattering matrices in effective 1D schemes, which proved not to fully capture the interplay between two-electron correlations and the realistic geometry of the device, as for electron bunching. Here, we privilege the exact solution by developing a scalable parallel numerical solver of the time-dependent Schr{\"o}dinger equation for two particles in a 2D realistic geometry. Moreover, the use of a two-particle Hamiltonian in our simulations allows us to easily include the \text{exact} Coulomb interaction of the two charges, in order to explore how electron-electron repulsion interplays with the exchange interaction of the fermionic system.

We initially describe, in Section II, our time-dependent numerical method to simulate two-particle transport in edge channels and the potential landscape we compute to reproduce the HOM interferometer. In Section III(A), we obtain the dynamical bunching probability in presence of exchange symmetry in the two-particle wave function, while in Section III(B) we study the effect of Coulomb repulsion between the two charges. Finally, in Section III(C) we present our measurement of the dynamical spatial entanglement between the two anti-bunched regions of the electronic HOM interferometer. In Section IV, we draw our conclusions.

\section{Physical system and Numerical model}\label{sec:physsys}
\subsection{Single-particle edge states as a basis}
\begin{figure}[t]
\includegraphics[width=1.\columnwidth]{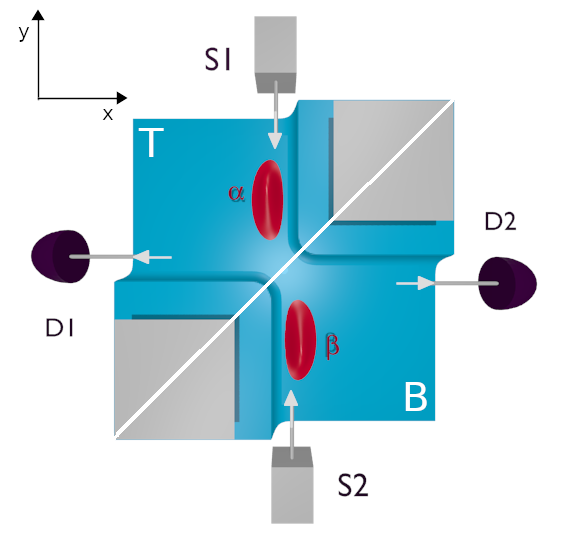}
\caption{Top-view of the HOM interferometer with the potential landscape reproducing the QPC (blue region) and the integrated single-particle density probability at $t=0$ (red wave packets). $\alpha$ and $\beta$ are the initial counterpropagating states for $\sigma=20$~nm, while $T$ and $B$ label the \textit{top} and \textit{bottom} regions we defined in the HOM (separated by the diagonal white line).}
\label{fig:topview}
\end{figure} 
We simulate the dynamics of two charges $-e$ with an effective mass $m^*$, that propagate in a 2DEG in the IQH regime. Here, a perpendicular magnetic field $\textbf{B} = (0,0,B)$~T is included in the single-particle Hamiltonian $H(x,y)$ by using the Landau gauge $\textbf{A}=(0,Bx,0)$. In presence of a confining lateral barrier $V(x)$, translationally-invariant in the $\hat{y}$-direction, the single-particle Hamiltonian is diagonalized by $\Psi_{n,k}(x,y)=e^{iky}\varphi_{n,k}(x)$, where the index $n$ refers to the Landau level, $k$ to the wave vector in the direction of propagation and $\varphi_{n,k}(x)$ diagonalizes the effective 1D Hamiltonian in the transverse direction: 
\begin{equation}\label{eq:H_effective}
H_{eff}(x)=-\frac{\hbar^2}{2m^*}\frac{\partial^2}{\partial x^2}+\frac{1}{2}m^*\omega_c^2(x-x_0(k))^2+V(x),
\end{equation}
with $x_0(k)=-\frac{\hbar k}{eB}$ and $\omega_c=\frac{eB}{m^*}$.
In proximity of the confining potential ($V(x)\neq 0$), the bending of the Landau levels generates conductive channels called edge states. Due to the chirality of the edge state $\varphi_{n,k}(x)$, an electron initialized in it can not be back-reflected by a potential roughness on its path, unless it is scattered by a narrow QPC to the counterpropagating state $\varphi_{n,-k}(x)$  at the opposite side of the confined 2DEG \cite{Venturelli2011_PRB, Beggi2015_JOPCM}.

In the electronic HOM experiment, particles are initialized in counterpropagating edge states at opposite sides of a beam splitter, as shown in Fig.~\ref{fig:topview}. In our geometry, only the first Landau level is filled, i.e. only $\varphi_{n=1,k}(x)$ is numerically computed from Eq.~(\ref{eq:H_effective}) by means of LAPACK routines. 
\begin{figure}[b]
\includegraphics[width=1.\columnwidth]{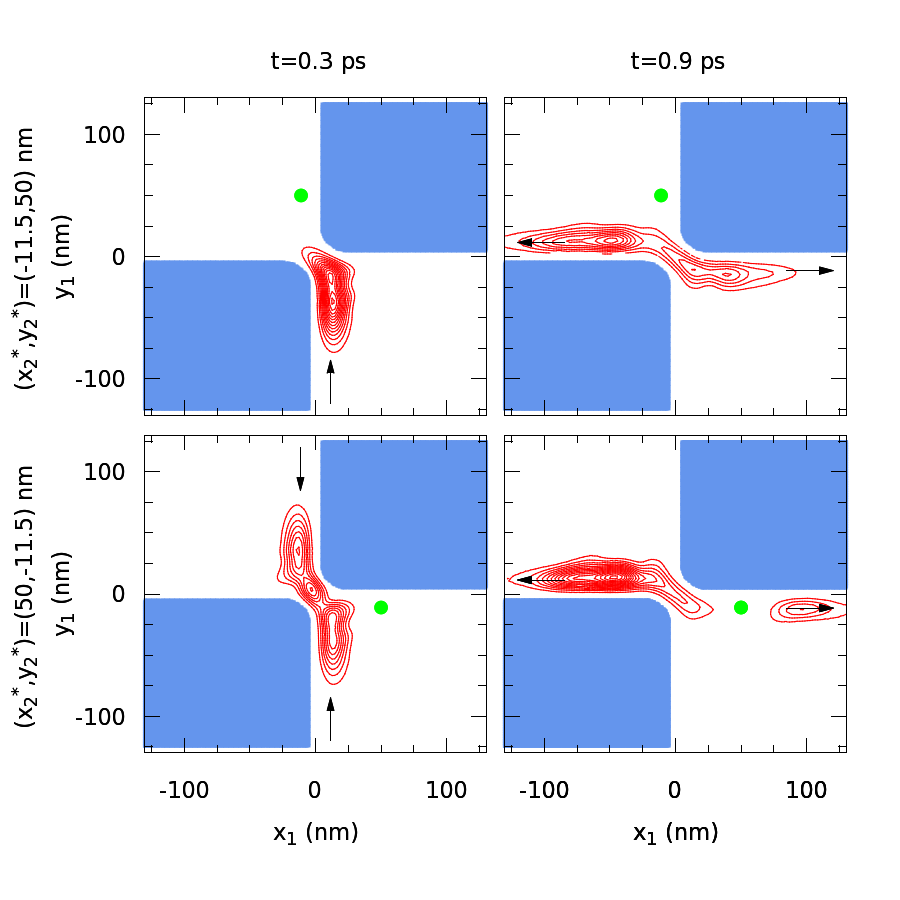}
\caption{Conditional probability (red contour lines) for particle 1 $P^*(x_1,y_1,x_2^*,y_2^*,t')=|\Psi(x_1,y_1,x_2^*,y_2^*,t')|^2$ at $t'=0.3$~ps (left column) and $t=0.9$~ps (right column) for the selected positions (green dots) of particle 2, $(x_2^*,y_2^*)=(-11.5,50)$~nm (first row) and $(x_2^*,y_2^*)=(50,-11.5)$~nm (second row), in the potential landscape of the HOM. Note that, due to Pauli exclusion principle, the wave function always vanishes at $(x_1,y_1)=(x_2,y_2)$. By selecting different couples of coordinates $(x_2^*,y_2^*)$ the conditional probability shows the evolution of one of the two or both single-particle wave packets, as indicated by the arrows.}\label{fig:pcond}
\end{figure} 

\subsection{Electron as a time-dependent superposition of single-particle edge states}
To include the energy dispersion of the electron state, edge states with different wave vector $k$ are combined, so that the single-particle wave packet reads
\begin{equation}\label{eq:singlepwp}
\psi(x,y)=\int dk F_{\sigma}(k,k_0,y_0)e^{iky}\varphi_{1,k}(x),
\end{equation}
where $F_{\sigma}(k,k_0,y_0)=\sqrt[4]{\sigma^2/2\pi^3}e^{-\sigma^2(k-k_0)^2}e^{-iky_0}$ is a Gaussian weight function centered at $k_0=-\frac{eB}{\hbar}x_0$ and with a dispersion $\sigma$ in the $\hat{y}$-direction. 
In order to avoid numerical errors induced by unrealistic step-like potentials, the wave packets are initialized next to confining barriers with a more realistic shape:
\begin{equation}\label{eq:pot_x}
V^{\alpha}(x)=V_b\mathcal{F_{\tau}}(x-x_b), \hspace{0.2cm}  V^{\beta}(x)=V_b\mathcal{F_{\tau}}(-x+x_b),
\end{equation}
where $\mathcal{F_{\tau}}(x)=(\exp(\tau x)+1)^{-1}$ is characterized by the smoothness $\tau$, height $V_b$, and $\alpha$ and $\beta$ label the two initialization regions of the device (Fig.~\ref{fig:topview}).
The indistinguishability of the two wave packets $\psi_{\alpha}(x,y)$ and $\psi_{\beta}(x,y)$ at $t=0$ is ensured by the equivalences $\tau^{\alpha}=\tau^{\beta}$ and $V_b^{\alpha}=V_b^{\beta}$, while the opposite direction of propagation is guaranteed by the the symmetry between the confining barriers of the 2DEG ($V^{\alpha}(x)=V^{\beta}(-x)$).   
The counterpropagating states $\psi_{\alpha}(x,y)$ and $\psi_{\beta}(x,y)$ are therefore characterized by an opposite wave vector of propagation $k_0^{\alpha}=-k_0^{\beta}$ and initial central position $y_0^{\alpha}=-y_0^{\beta}$, so that they impinge on the beam splitter simultaneously. Numerical values used in our simulations are $V_b=10$~eV and $\tau=3$~nm.

\subsection{Numerical solution of the two-particle Schr\"odinger equation in four spatial dimensions plus time.}
By assuming a symmetric spin part of the wave function, the Slater determinant is finally computed from the orbital states $\psi_{\alpha}(x,y)$ and $\psi_{\beta}(x,y)$, so that exchange symmetry is included in the two-electron state:
\begin{eqnarray}\label{eq:twopstate}
\Psi=\frac{\psi_{\alpha}(x_1,y_1)\psi_{\beta}(x_2,y_2)-\psi_{\alpha}(x_2,y_2)\psi_{\beta}(x_1,y_1)}{\sqrt{2}}.
\end{eqnarray}

\begin{figure}[b]
\includegraphics[width=1.\columnwidth]{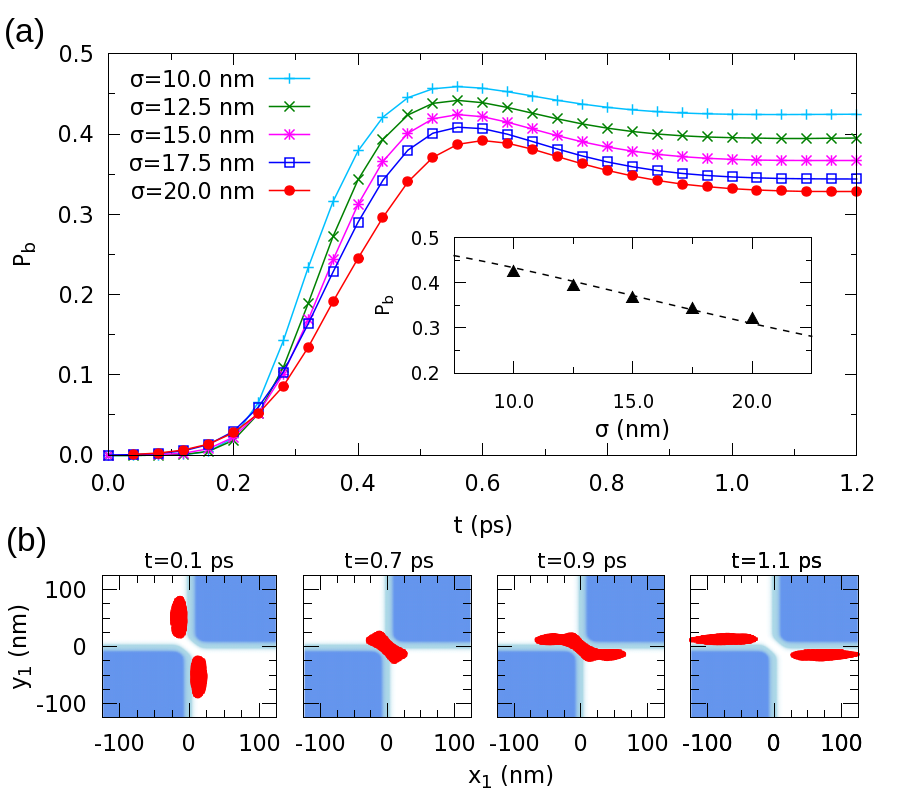}
\caption{Numerical simulations of the exchange-symmetry driven HOM interferometer. (a) Bunching probability of the two-fermion state in absence of Coulomb interaction for different spatial dispersions $\sigma$. (inset) Numerical fit of Eq.~(\ref{eq:bunching}) (black dashed line) with the stationary bunching probabilities in Fig.\ref{fig:bunchingexchange}(a) at $t=1.2$~ps ($\blacktriangle$). This provides the geometrical parameter in Eq.(\ref{eq:rtqpc}) for our QPC, $a=60\pm1$~nm. (b) Unconditional probability of particle 1 $P(x_1,y_1,t')=\int\int dx_2dy_2|\Psi(x_1,y_1,x_2,y_2,t')|^2$ at different time steps $t'$ in the potential landscape of the HOM. The snapshots show that the stationary regime is achieved between $0.9$~ps and $1.2$~ps. Note that due to exchange correlation, the unconditional probability of particle 1 shows the evolution of both wave packets.}
\label{fig:bunchingexchange}
\end{figure}
Together with accounting for the realistic geometry of the device, the explicit computation of the two particle wave function in the 4D configuration space allows one to exactly reproduce the system dynamics in presence of electron-electron interaction $V_{12}(x_1,y_1,x_2,y_2)$. However, the memory burden to allocate the 4D wave function increases the computational cost with respect to the single-electron version of our software  \citep{Bellentani2018_PRB,Beggi2015_JOPCM}. The present numerical simulations entail indeed distributed-memory parallelization techniques \cite{SURBOOK2006} and the support of high-performance computing facilities; to face the problem of memory allocation, we partition the $(x_2,y_2)$ domain of the configuration space between interconnected supercomputing nodes with the MPI paradigm, while a parallel version of the Split-Step Fourier method  \citep{Kramer2010_PS} evolves the distributed two-particle wave function.

The latter time-dependent algorithm has already been validated to study electron transport in single-qubit Hall interferometers in Ref.~\onlinecite{Bellentani2018_PRB} and Ref.~\onlinecite{Beggi2015_JOPCM}. Here, we use its extension to the two-particle case. In the Landau gauge, $\textbf{A}_i=(0,Bx_i,0)$ with $i=1,2$, the two-particle Hamiltonian
\begin{equation}
\hat{H}_{12}=\frac{(\hat{p}_1-q\hat{A}_1)^2}{2m^*}+\frac{(\hat{p}_2-q\hat{A}_2)^2}{2m^*}+\hat{V}_1+\hat{V}_2+\hat{V}_{12},
\end{equation} is split by using Trotter-Suzuky factorization and included in the the evolution operator $\hat{U}(t,0)$ for a total evolution time $t=N\delta t$: 
\begin{align}\label{eq:evoloperat}
\hat{U}(t,0)=&[e^{-\frac{i}{\hbar}\delta t\cdot (\hat{V}_1+\hat{V}_2+\hat{V}_{12})}\textsf{F}^{-1}_{y_1,y_2}e^{-\frac{i}{\hbar}\delta t\cdot\hat{T}_{y_1,y_2}} \nonumber \\
&\textsf{F}_{y_1,y_2}\textsf{F}_{x_1,x_2}^{-1}e^{-\frac{i}{\hbar}\delta t\cdot \hat{T}_{x_1,x_2}}\textsf{F}_{x_1,x_2}]^N.
\end{align}
$\textsf{F}_{y_1,y_2}$ ($\textsf{F}^{-1}_{y_1,y_2}$) and $\textsf{F}_{x_1,x_2}$ ($\textsf{F}^{-1}_{x_1,x_2}$) are parallel 2D Fourier transforms (antitrasforms) performed on the single-particle real-space coordinates $x_i,y_i$ ($i=1,2$). Fourier transforms exploit the locality of the modified kinetic operators $\hat{T}_{x_1,x_2}$ ($\hat{T}_{y_1,y_2}$) in the reciprocal space $[k_{x_1},k_{x_2}]$ ($[k_{y_1},k_{y_2}]$), while the single-particle $V_{1},V_{2}$ and two-particle $V_{12}$ potential operators are computed \textit{exactly} in the real-space. 

The two-particle evolution in the HOM experiment is then affected by the presence of the beam splitter, to partition the impinging wave packets in a reflected and transmitted component with the same probability in ideal conditions. In this geometry, the beam splitter is realized by a QPC, i.e. a narrow constriction in the potential confining barrier, that partially scatters the traveling charge to the counterpropagating edge channel.    
We design the QPC by using potential barriers as in Eq.~(\ref{eq:pot_x}), so that the potential profile reads 
\begin{eqnarray}\label{eq:pot_xy}
V(x,y)=&V_b\mathcal{F}(-x-\infty)\mathcal{F}(x-x_L)\mathcal{F}(-y-\infty)\mathcal{F}(y-y_B) \nonumber \\ 
+&V_b\mathcal{F}(-x+x_R)\mathcal{F}(x-\infty)\mathcal{F}(-y+y_T)\mathcal{F}(y-\infty) \nonumber \\
\end{eqnarray}
where $V_b$ is the height of the confining potential, $x_L$, $x_R$ the left and right side of the barrier along the $\hat{x}$-direction, and $y_T$ and $y_B$ the top and bottom side of the barrier along the $\hat{y}$-direction, respectively. Such parameters of the single-electron potential $V(x,y)$ in Eq.~(\ref{eq:pot_xy}) have been tuned to transmit and reflect the impinging single-electron wave packet in Eq.~(\ref{eq:singlepwp}) with $50\%$ probability, as achieved in the Mach-Zehnder interferometer of Ref.~\onlinecite{Beggi2015_JOPCM}. 
In the present geometry, the electron beam splitter has a symmetric opening about $30$~nm. 
Moreover, our choice of the smoothness parameter $\tau$ of Eq.~(\ref{eq:pot_x}) and the transverse position $x_0$ lead to a group velocity\cite{Beggi2015_JOPCM} of about $150$~nm/ps for our wave packets.

Ref.~\onlinecite{Beggi2015_JOPCM} shows the operability of the QPC in the IQH regime and the consequence of the distinctive Fermi-like energy selectivity of the present beam splitter. In this geometry, if a proper initial position $x_0(k_0)$ of the wave packet ensures an integrated transmission probability of about $50\%$ at the QPC, the Fermi-like energy selectivity of the partitioner induces not-complete overlap between the transmitted and reflected components of the single-electron wave packet in the $k_y$ space. This feature not only affects the visibility in early implementations of the Mach-Zehnder interferometer, but, as shown in the next section, it also has a key role in the apparent violation of the Pauli exclusion principle in the HOM experiment \cite{Marian2015_JPCM}.

\section{RESULTS}\label{sec:res}
\begin{figure}[t]
\includegraphics[width=1.\columnwidth]{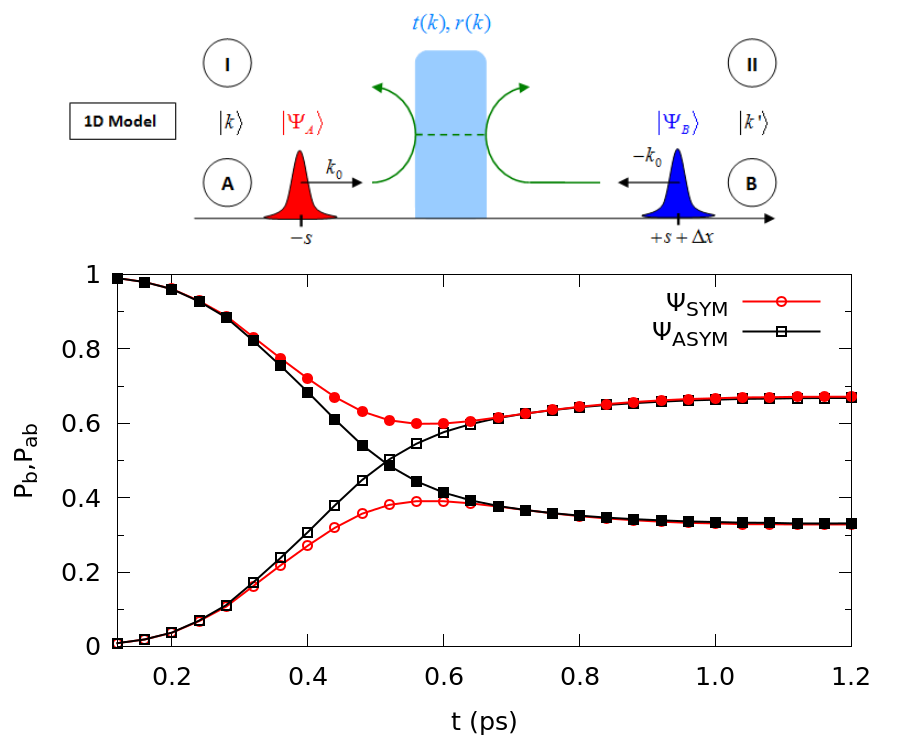}
\caption{(Top) 1D model for the two-particle scattering at the QPC: two wave packets of plane waves with opposite central momentum impinge on a 1D potential barrier. (Bottom) Bunching (empty dots) and antibunching (full dots) probabilities for an antisymmetric (red) and symmetric (black) two-particle wave function with $\sigma=20$~nm. Only exchange symmetry is present.}
\label{fig:1Dmodel}
\end{figure}
In the following, we present our exact numerical simulations of the HOM experiment in the IQH regime, with $B=5$~T. By considering the typical parameters of GaAs ($m^*=0.067m_e$) for the hosting material, we tune the spatial broadening of the two indistinguishable electron wave packets ($\sigma=10,12.5,15,17.5$ and $20$~nm), and we observe the interplay between the HOM geometry and two-electron correlations in different scenarios, where the exchange and/or Coulomb interaction are included. 

The dynamical properties of the system are completely determined from the 4D wave function $\Psi(x_1,y_1,x_2,y_2,t)$, which is iteratively computed at each time step. As an example, Fig.~\ref{fig:pcond} displays the conditional probability of one particle by selecting the coordinates of the other at two different positions and at two different times during the evolution in the HOM interferometer. We stress that the huge computational load of the 4D wave function requires to reduce the number of degrees of freedom in order to produce viable information, as  to measure the dynamical bunching (antibunching) probability $P_{b}$ ($P_{ab}$), which corresponds to the joint probability of finding the two particle in the same output (opposite outputs)~\citep{Marian2015_JPCM}. The real-space domain $(x,y)$ is divided in two regions, labeled \textit{top} ($T$) and \textit{bottom} ($B$) as shown in Fig.~\ref{fig:topview}, so that the configuration space $[x_1,y_1,x_2,y_2]$ is partitioned into 4 distinguishable regions labeled $S_{i,j}$, where the indexes $i,j=T,B$ refer to the first and to the second particle respectively. The joint probability of detection at the same side of the device is
\begin{eqnarray}\label{eq:p_bunching}
P_{b}&=&\int_{S_{TT}} |\Psi|^2dx_1dy_1dx_2dy_2 \nonumber \\ &+& \int_{S_{BB}} |\Psi|^2dx_1dy_1dx_2dy_2, 
\end{eqnarray}
while the antibunching probability is similarly computed by integrating over $S_{TB}$ and $S_{BT}$; bunching and antibunching probability are related by $P_{ab}=1-P_{b}$. 

\subsection{Effect of QPC scattering asymmetry}\label{subsec:resen}
In a steady-state framework, where the injection of plane waves is assumed, $P_{b}$ is expected to be zero for an half-reflecting beam splitter, due to the exchange symmetry of a fermionic two-particle state $\Psi$, even neglecting Coulomb interaction  \citep{Buttiker1992_PRL}.
Differently, by using a time-dependent model in a 1D effective framework, it is proved  \citep{Marian2015_JPCM} that the energy broadening of the single-electron state affects the bunching probability. This behavior is confirmed by our full-scale time-dependent simulations of the two-electron bunching in 2D for different wave packet sizes, reported in Fig.~\ref{fig:bunchingexchange}. Here, in presence of exchange symmetry only, the  bunching probability of the two-fermion wave function initially varies with time, reaching its maximum  when the two particles interact at the QPC. Then, the two charges leave the QPC only partially from different outputs, so that the bunching probability decreases without vanishing. As shown by comparing $P_b(t)$ to the evolution of the two-particle probability (bottom panel of Fig.~\ref{fig:bunchingexchange}), the stationary regime is achieved at $t\simeq 1.0$~ps. By increasing $\sigma$, the final bunching probability decreases linearly, so that we expect that the full antibunching is restored in the plane wave limit, i.e. $\sigma\rightarrow \infty$. 

The stationary trend of $P_b(\sigma)$ in the exclusive presence of exchange symmetry can be estimated analytically by means of the one-dimensional effective model we already validated to study single-qubit interferometers \citep{Bellentani2018_PRB,Beggi2015_JOPCM}. The model relies on the chirality of edge states to map our two-dimensional system in the IQH regime with wave packets of plane waves impinging on a 1D barrier, as schematically depicted in the top panel of Fig.~\ref{fig:1Dmodel}. The edge state $\varphi_{n=1,k}(x)e^{iky}$ is replaced by a plane-wave $|k\rangle$, with an effective parabolic dispersion $E(k)$ characterized by a magnetic mass $m_B^*$. 
In the reciprocal space, the single-particle wave functions are defined on the pseudo-spin basis $\{ |k_\alpha\rangle,|k_\beta\rangle \}$, where $|k_\alpha\rangle=-|k_\beta\rangle=|k\rangle$. Neglecting Coulomb interaction, the effect of the potential barrier is described by a single-particle scattering matrix:
\begin{equation}
\hat{S}=\begin{pmatrix}
r(k) & t(k) \\
t(k) & r(k)
\end{pmatrix}
\end{equation}  
where $r(k)$ and $t(k)$ are empirical equations that model \citep{Beggi2015_JOPCM} the energy selectivity of a realistic QPC:
\begin{equation}
\Big[ \begin{matrix}\label{eq:rtqpc}
r(k) \\ t(k)
\end{matrix} \Big]= \Big[ \begin{matrix}
1 \\ i
\end{matrix} \Big] \exp\big(-\frac{(\mp a(k-k_0)+\gamma)^2}{8\gamma}\big).
\end{equation}  
The $a$ parameter depends on the smoothness $\tau$ and height $V_b$ of the QPC in Eq.~(\ref{eq:pot_xy}), while $\gamma=4 \, \text{ln} \, 2$ in our model. The scattering at the QPC splits the single-particle wave packet $|\psi_{\alpha(\beta)}\rangle$ in two contributions, that we label $|\alpha(\beta),R\rangle$ for the reflected and $|\alpha(\beta),T\rangle$ for the transmitted component:
\begin{eqnarray}\label{eq:trwp}
|\alpha(\beta),R\rangle = \int dk F_{\sigma}(k,k_0^{\alpha(\beta)},y_0^{\alpha(\beta)})r(k)|k\rangle, \\
|\alpha(\beta),T\rangle = \int dk F_{\sigma}(k,k_0^{\alpha(\beta)},y_0^{\alpha(\beta)})t(k)|k\rangle.
\end{eqnarray}
The scattered single-particle wave packets $|\psi_{\alpha(\beta)}\rangle'=|\alpha(\beta),R\rangle + |\alpha(\beta),T\rangle$ are inserted in Eq.~(\ref{eq:twopstate}) to compute the bunched state:
\begin{equation}
|\psi_{bun}\rangle =\frac{|\alpha R\rangle|\beta T\rangle+|\alpha T\rangle|\beta R\rangle -|\beta R\rangle|\alpha T\rangle-|\beta T\rangle|\alpha R\rangle}{\sqrt{2}},
\end{equation}
where the first ket refers to particle 1, while the second to particle 2.
Taking into account the orthogonality of $|\alpha(\beta)R\rangle$ and $|\alpha(\beta)T\rangle$ due to their spatial separation, the bunching probability $P_{b}=\langle \psi_{bun} | \psi_{bun} \rangle$ is calculated:
\begin{eqnarray}\label{eq:bunchingket}
P_b&=\langle\alpha R|\alpha R\rangle \langle \beta T | \beta T\rangle +\langle \alpha T|\alpha T\rangle \langle \beta R|\beta R \rangle \nonumber \\ &-2|\langle \alpha T | \beta R \rangle |^2.  
\end{eqnarray} 
We assume the ideal condition $\langle \alpha (\beta)R | \alpha (\beta)R\rangle = \langle \alpha (\beta)T | \alpha (\beta)T\rangle =1/2 $, while the overlap integrals $\langle \alpha (\beta)R | \beta(\alpha)T \rangle$ are computed exploiting Eq.~(\ref{eq:rtqpc}). In case of asymmetry between the initial positions of the two wave packets ($y_0^{\alpha}=y_0^{\beta}+\Delta y$), the bunching probability for a QPC with $\Sigma^2=\sigma^2+a^2/8\gamma$ reads: 
\begin{equation}\label{eq:bunching}
P_{b}=\frac{1}{2}-\frac{1}{2}\frac{\sigma^2}{\Sigma^2}e^{\frac{-\Delta y^2}{4\Sigma^2}}.
\end{equation} 

In the inset of Fig.~\ref{fig:bunchingexchange} (top panel), the exact bunching probabilities computed numerically at $t=1.2$~ps are fit by means of Eq.~(\ref{eq:bunching}) with $\Delta y=0$.  This 1D effective model not only confirms quantitatively the results of our 2D simulations, but it also explains how the energy broadening affects the Pauli dip. 
According to Eq.~(\ref{eq:rtqpc}), the transmitted and reflected components of the single-particle scattered wave packets are centered at different wave vectors, thus decreasing the overlap $\langle \alpha T | \beta R \rangle$ in Eq.~(\ref{eq:bunchingket}). 
Two strategies are therefore possible to reduce $P_b$: the manipulation of the spatial broadening $\sigma$, as shown by Fig.~\ref{fig:bunchingexchange}, or a variation in the smoothness of the QPC. The former case is realized when $\sigma\rightarrow \infty$ (plane-wave limit), while the latter if $a\rightarrow 0$ (flat energy selectivity) in Eq.~(\ref{eq:p_bunching}). Regarding the smoothing of the energy selectivity, a different geometry, as the beam splitter at bulk filling factor 2 in Ref.~\onlinecite{Bellentani2018_PRB}, represents a possible solution.

We additionally remark that in Eq.~(\ref{eq:bunchingket}) the overlap $-2|\langle \alpha T | \beta R \rangle|^2$ decreases the bunching probability due to the antisymmetry of the two-electron state. In case of a symmetric wave function the same term is expected to be positive ($+2|\langle \alpha T | \beta R \rangle|^2$), as shown in the bottom panel of Fig.~\ref{fig:1Dmodel}. Here, we compare the dynamical bunching and antibunching probabilities of a symmetric/antisymmetric two-particle wave function that propagates in our geometry of the HOM interferometer: in the stationary regime, the two configurations are characterized by exchanged values of $P_b$ and $P_{ab}$.  
\begin{figure}[t]
\includegraphics[width=1\columnwidth]{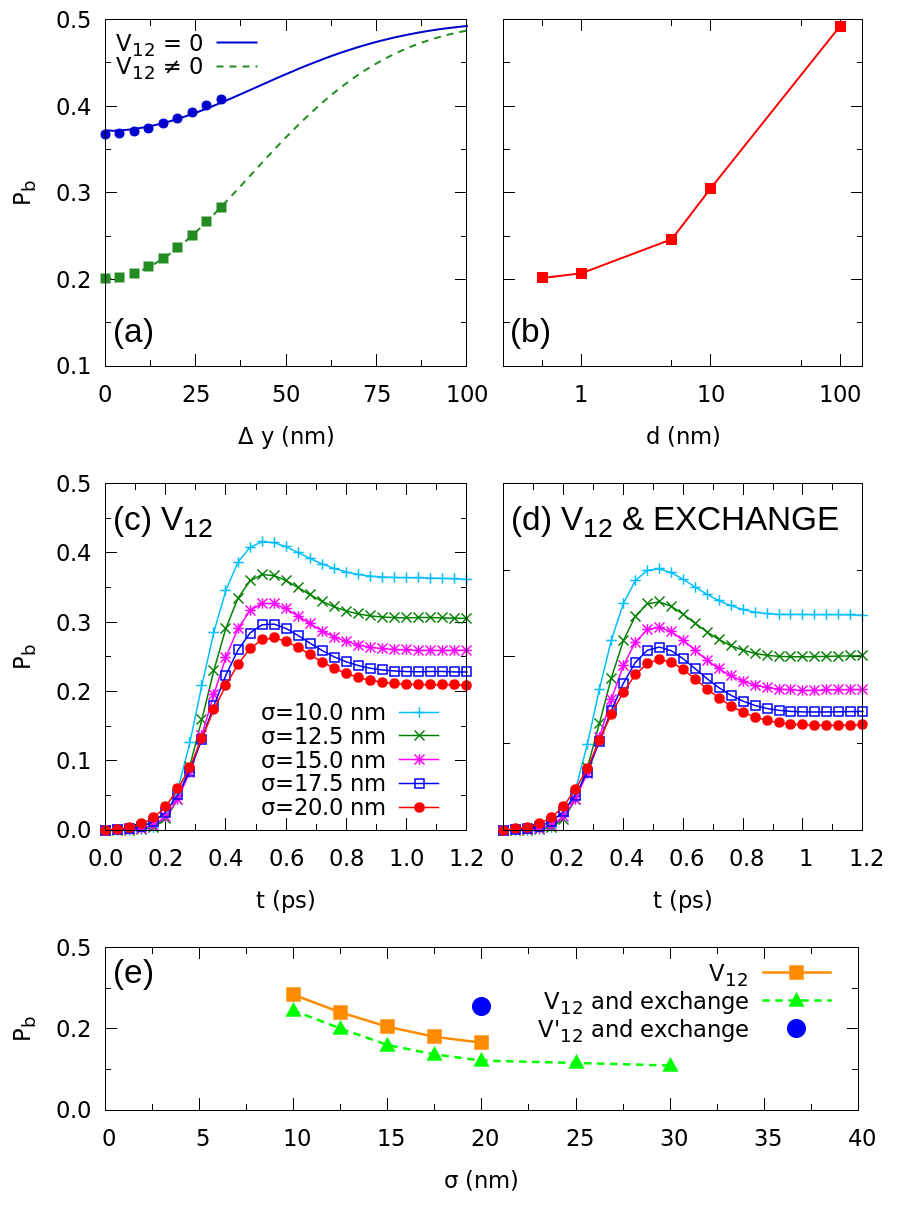}
\caption{(a) Bunching probability as a function of the initial displacement $\Delta y$ between two indistinguishable wave packets ($\sigma=15$~nm) with zero (blue) and non zero (green) Coulomb interaction $V_{12}$ with $d=1$~nm. For $V_{12}=0$ the numerical data (blue dots) are compared to Eq.~(\ref{eq:bunching}) (blue line) with $\alpha=60$~nm, while for $V_{12}\neq 0$ numerical data (green dots) are fit by the equation $g(x)$ (green dashed line) as explained in the main text. The fit provides an effective broadening $\sigma_{eff}=21.75\pm0.03$~nm and an effective geometrical parameter $a_{eff}=84.0\pm0.1$~nm. (b) Stationary bunching probability ($t=1.2$~ps) in presence of long-range Coulomb interaction without exchange symmetry for different $d$ parameters in Eq.~(\ref{eq:coulomb}) and $\sigma=20$~nm . (c) Bunching probability in presence of long-range Coulomb interaction without exchange symmetry and (d) with exchange symmetry for different $\sigma$ and $d=1$~nm. (e) Comparison between the stationary bunching probability of two distinguishable electrons (yellow line) and indistinguishable electrons (green line) with long-range Coulomb interaction. The stationary bunching probability of two indistinguishable electrons in presence of a screened Coulomb interaction is also reported (blue circle) for the case $\sigma=20$~nm and a damping $\sigma_c=5$~nm, see Eq.~(\ref{eq:screenedCoulomb}).}
\label{fig:buncoulomb} 
\end{figure}
Differently, as indicated by Eq.~(\ref{eq:bunching}), the symmetry of the two-particle wave function is not expected to affect $P_b$ at $\Delta y \rightarrow \infty$. The stationary bunching probability for different initial displacements $\Delta y$ and $\sigma=15$~nm is reported in Fig.\ref{fig:buncoulomb}(a) (blue dots for $V_{12}=0$). In our time-dependent scenario, a spatial mismatch in the initial position of the two electron states ($y_0^{\alpha}-y_0^{\beta}=\Delta y\neq 0$) modifies the time at which each wave packet is expected to impinge on the QPC, so that the overlap between the two wave packets decreases, as well as the effect of their exchange symmetry. The classical probability of joint detection for two distinguishable particles is gradually restored with $\Delta y \rightarrow \pm\infty$. Eq.~(\ref{eq:bunching}) shows that the characteristic length for this process is $\lambda=2\sqrt{\sigma^2+\frac{a^2}{8\gamma}}\approx 30$~nm for $\sigma=15$~nm, as confirmed by the fitting curve $P_b(\Delta y)$ (blue solid line) in Fig.\ref{fig:buncoulomb}(a), which vanishes at $3\lambda\simeq 90$~nm.

\subsection{Effect of Coulomb interaction}\label{subsec:rescou} 
If our single-particle approach provides enough information to understand electron bunching probability in presence of exchange-symmetry only, the introduction of Coulomb interaction requires a full-scale two-particle approach. We introduce long-range electron-electron repulsion between the two carriers by adding to the two-particle Hamiltonian the potential
\begin{equation}\label{eq:coulomb}
V_{12}(x_1,y_1,x_2,y_2)=\frac{e^2}{4\pi\epsilon\sqrt{(x_1-x_2)^2+(y_1-y_2)^2+d^2}},
\end{equation}
where $\epsilon$ is the medium permittivity and $d$ accounts for the finite thickness of the 2D system in the divergence at $x_1=x_2$ and $y_1=y_2$.

To evaluate the interplay between exchange symmetry and Coulomb interaction, we initially compute the bunching probability for a product state $\Psi_C=\psi_\alpha(x_1,y_1)\psi_\beta(x_2,y_2)$, where the antibunching is exclusively generated by electron-electron repulsion. In Fig.~\ref{fig:buncoulomb}(b) we report the bunching probability at final time $t=1.2$~ps with $\sigma=20$~nm for different values of the $d$ parameter. Note that, for a Coulomb repulsion small enough ($d=100$~nm), $P_b$ reaches the limit $1/2$. Indeed, in absence of the exchange symmetry, the two wave packets evolve independently and the bunching probability in Eq.~(\ref{eq:bunchingket}) does not contain the overlap $\langle \alpha T|\beta R\rangle$. $P_b$ equals $1/2$ only if $\langle \alpha(\beta)T|\alpha(\beta)T\rangle=\langle \alpha(\beta)R|\alpha(\beta)R\rangle=1/2$, namely the two initial wave packets are properly initialized, so that they are transmitted with $50\%$ probability by the QPC, as in the present case. Additionally, Fig.~\ref{fig:buncoulomb}(b) shows that the Coulomb-driven scattering at the QPC does not produce perfect antibunching when $d\rightarrow 0$, but rather $P_b$ saturates by decreasing the $d$ parameter. This differs from a one-dimensional system, where the two electron are confined on the same rail, e.g. in the $\hat{x}$-direction. In the latter case, they are forced to get across $x_1=x_2$, so that they feel an effective infinite barrier during the scattering for $d=0$. In the two-dimensional geometry, alternative paths with a finite barrier are present, and each charge is only partially reflected by the Coulomb potential for $d=0$, producing partial bunching.  

Fig.~\ref{fig:buncoulomb}(c) shows the evolution of $P_b(t)$ for the separable case with different values of $\sigma$ at $d=1$~nm. Finally, we add the exchange symmetry to the interacting system (Fig.~\ref{fig:buncoulomb}(d)). In our operating regime, the Coulomb repulsion dominates on the exchange interaction, which only shifts the bunching probability to lower values. 
We also show, in Fig.~\ref{fig:buncoulomb}(a), the stationary bunching probability of indistinguishable and interacting electrons for different initial displacements $\Delta y$ and $\sigma=15$~nm: numerical data (green squares) are fit by $g(x)=\frac{1}{2}\cdot[1-\frac{\sigma_{eff}^2}{\sigma_{eff}^2+\alpha_{eff}^2/8\gamma}\exp(-\frac{x^2}{4(\sigma_{eff}^2+a_{eff}^2/8\gamma)})]$ (green dashed line), which corresponds to Eq.~(\ref{eq:bunching}) with an effective $\sigma_{eff}$ and $a_{eff}=2\sqrt{2\gamma}\sqrt{\Sigma_{eff}^2-\sigma_{eff}^2}$ used as fitting parameters. The numerical fit provides $\sigma_{eff}\simeq 21$~nm and $a_{eff}\simeq 86$~nm, that are larger than the actual ones (Fig.\ref{fig:bunchingexchange}(a), inset) without Coulomb interaction. 

As visible in both separable (Fig.~\ref{fig:buncoulomb}(c)) and non separable (Fig.~\ref{fig:buncoulomb}(d)) interacting scenarios, the effect of long-range Coulomb interaction turns out to depend on the spatial broadening of the wave packet.
The stationary bunching probabilities of the separable and non-separable cases are displayed in Fig.~\ref{fig:buncoulomb}(e). In presence of long-range Coulomb interaction, $P_b(\sigma)$ clearly differs from the almost-linear one in the inset of Fig.~3(a), where the antibunching is exclusively driven by the exchange interaction. Two additional numerical simulations for indistinguishable electrons with $\sigma=25$ and $30$~nm confirm that, in our operating regime, the bunching probability saturates to a non zero value for larger wave packets. 

Finally, we present an additional simulation where the interaction between two indistinguishable electrons with $\sigma=20$~nm is screened by an exponential damping:
\begin{equation} \label{eq:screenedCoulomb}
V'_{12}(x_1,y_1,x_2,y_2)=\frac{Ce^2}{4\pi\epsilon}\frac{e^{-\frac{\sqrt{(x_1-x_2)^2+(y_1-y_2)^2}}{\sigma_c}}}{\sqrt{(x_1-x_2)^2+(y_1-y_2)^2+d^2}},
\end{equation} 
as done in Ref.~\onlinecite{Marian2015_JPCM} for an effective 1D geometry. 
The parameter $C$ is a global constant that quantifies the interaction, and $\sigma_c$ determines its spatial range. 
Fig.~\ref{fig:buncoulomb}(e) reports the stationary bunching probability compared to the case of unscreened Coulomb interaction. 
The numerical value for the screened case is almost identical to the bunching probability in presence of exchange alone, reported in the inset of Fig.3(a). 
This suggests that, for the present damping parameters ($C=1$ and $\sigma_c=5$~nm), the regime of exchange-driven bunching is restored, and the dominance of the Coulomb interaction is gradually suppressed in the intermediate regimes.

\subsection{Spatial entanglement}\label{subsec:ent}
We finally provide a dynamical estimation of the spatial entanglement for the antibunched configuration in presence of exchange and/or Coulomb interaction. 
\begin{figure}[b]
\includegraphics[width=1\columnwidth]{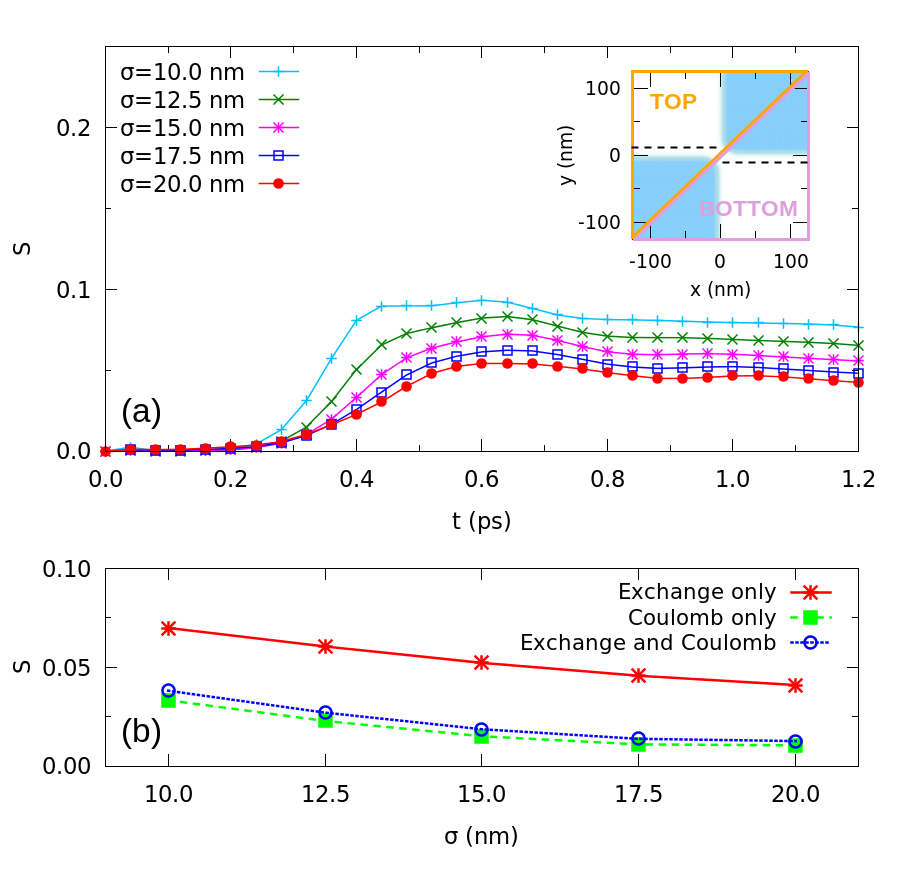}
\caption{Entanglement between $T$ and $B$ regions in the antibunched configuration. (a) Von Neumann entropy for different $\sigma$ and $d=1$~nm in presence of exchange symmetry only; the dotted black lines in the inset show where the full scale wave function is projected with respect to the top-view of the potential profile. (Bottom) Comparison between the stationary Von Neumann entropy in the different scenarios. }
\label{fig:entanglement}
\end{figure}
The Hilbert space is divided into two separate subsystems, the \textit{top} ($T$) and the \textit{bottom} ($B$) region in Fig.~\ref{fig:entanglement}, so that $H=H_T\otimes H_B$. 
Due to the indistinguishability of the two electrons, entanglement is measured by assuming particle 1 in the top region ($(x_1,y_1)\in T$), and particle 2 in the bottom one ($(x_2,y_2)\in B$), while the opposite configuration is equally entangled due to the symmetry of the two wave packets. To reduce the computational burden, we exploit the chiral nature of edge channels by projecting the two-particle wave function on 1D slices of the subsystem: $y_1$ and $y_2$ coordinates are fixed at the expected maxima of the scattered single-particle wave packet in $T$ and $B$ regions, in the present case $y_1^*=11.5$~nm and $y_2^*=-11.5$~nm (black dashed lines in the inset of Fig.~\ref{fig:entanglement}). 

The conditional two-particle wave function, $\phi(x_1,x_2)=\Psi(x_1,y_1^*,x_2,y_2^*)$ is renormalized, and the density matrix reads:
\begin{equation}
\rho_{TB}(x_1,x_2;x_1',x_2')=\phi(x_1,x_2)\phi^*(x_1',x_2'),
\end{equation}
whose Von Neumann entropy does not depend on the subspace chosen to be traced out. Therefore, we compute the reduced density matrix
\begin{equation}\label{eq:reduced_density}
\rho_T(x_1,x_1')=\int_{x_2\in B}dx_2 \ \phi(x_1,x_2)\phi^*(x_2,x_1'),
\end{equation}
on the lattice points of our domain B. We calculate the spatial entanglement by means of the Von Neumann entropy
\begin{equation}
S=Tr[\rho_T \ln(\rho_T)].
\end{equation}

Fig.~\ref{fig:entanglement}(a) shows the dynamics of the Von Neumann entropy of the wave-function spatial distribution in the $T$ and $B$ regions with exchange symmetry only in the antibunching configuration. The entanglement is quenched by an increase of the spatial distribution of the wave packet, so that we expect it to vanish in the plane-wave limit. In Fig.~\ref{fig:entanglement}(b) we compare the stationary Von Neumann entropy for the cases of: (i) exchange symmetry alone (red solid line), (ii) a separable wave function with Coulomb interaction (green dashed line), (iii) Coulomb interaction and symmetric wave function (blue dotted line). In the latter, the Coulomb repulsion, which acts as an additional barrier, further prevents the two particles to reach the opposite regions thus damping entanglement. The Von Neumann entropy in the interacting scenario with exchange interaction does not differ from the one in the distinguishable case: the two electrons are already prevented to occupy the same coordinates due to the infinite barrier represented by $V_{12}$ at $(x_1,y_1)=(x_2,y_2)$.     
We stress that the entanglement assessed in Fig.~\ref{fig:entanglement} represents the degree of non-separability of the spatial representation of the two-particle wave function and its possible exploitation as a resource for quantum information processing is not straightforward and beyond the scope of this paper.

\section{Conclusions}\label{sec:conc}
We realized a full-scale numerical simulation of an electronic HOM interferometer in the IQH regime that contributes to shed light on the apparent violation of Pauli exclusion principle in two-electron bunching, by including exactly the interplay between a realistic geometry of the QPC in 2D and the exchange and correlation of the two-particle wave function. 
A full understanding of such interplay is important for the implementation of a HOM device for quantum computing protocols, e.g. to measure the degree of indistinguishability of electrons ge\-ne\-ra\-ted from different sources~\citep{Kataoka2016_PRL}. 
Moreover, from an experimental perspective, the access to the dynamics of the full two-particle wave function and its dependence on several parameters, as the spatial dispersion of the carriers, represents a formidable ingredient to assess the origin of low-frequency fluctuations in the electrical current.  In fact, the low-frequency noise is proportional to the overlap between the two electron states, and provides the degree of indistinguishability of the two electrons impinging on the beam splitter. By introducing desynchronization between the two sources the full minimum of the Pauli dip can be characterized without resorting to a challenging detection of coincidence counts.

In contrast to traditional approaches in literature, that are based on 1D effective models implemented in stationary frameworks and with effective scattering matrices for the QPC, we privilege the exact solution of the 4D time-dependent Schr\"odinger equation for two particles in a 2D real space. 
The exact solution requires a parallel implementation of our numerical solver by means of the MPI library. Thanks to this effort, we provide dynamical measurements of the bunching probability for a two-particle system, where the electrons are initialized in Gaussian wave packets of the \textit{exact} edge states confined at the barrier, coherently to the most recent single-electron injection protocols in Hall nanodevices. 
The time evolution of the two interacting and correlated wave packets successfully reproduce the two-electron dynamics in the present interferometer. The decrease in the bunching probability by increasing the spatial localization confirms, for a full scale 2D spatial geometry, the findings of Ref.~\onlinecite{Marian2015_JPCM}, where this effect is explained with a 1D time-dependent model as a signature of the non-orthogonality between the states scattered by the potential barrier. 

In addition to the full-scale numerical simulation, we provide a simplified analytical model to relate the stationary bunching probability to the non-perfect overlap between the transmitted and reflected wave-packets from the QPC. This model clarifies the interplay between the spatial dispersion of the wave packet $\sigma$ and the geometry of the QPC, which is encoded in the single-particle parameter $\Sigma$ of Eq.~(\ref{eq:bunching}). 

We show how the perfect antibunching is recovered in the plane-wave limit and point out the role of exchange symmetry by simulating the HOM experiment both for a symmetric and an antisymmetric wave function. As an additional advantage in treating exactly the two-particle scattering, we include electron-electron repulsion in the Hamiltonian to evaluate the interplay with the fermionic statistics. We observe that in a 2D real-space geometry, differently from the typical 1D scenario adopted in literature, the bunching probability does not vanish for an infinite repulsive Coulomb interaction. 
Additionally, for unscreened interacting particles, we find that $P_b(\sigma)$ saturates to non zero values. Our conclusions do not contradict the results obtained in Ref.\onlinecite{Bocquillon2013_S}, as in the experiment the device does not generate strongly-localized excitations, but rather wave packets with an emission time of the order of tens of picoseconds, i.e. 2 orders of magnitude larger than in our geometry. Furthermore, by including an exponentially-decaying screening in our interacting regime, we show how the effect of Coulomb repulsion can be suppressed with a proper choice of the damping length, so that the limit of exchange-driven bunching is restored, also for interacting particles.

Finally, a dynamical measurement of the spatial Von Neumann entropy between the top and bottom regions of the device allows us to assess the spatial entanglement between the two antibunched carriers; we found that long-range Coulomb interaction quenches the entanglement by enhancing the Pauli dip with respect to the antisymmetry alone. 
This study represents the starting point for the simulation of more sophisticated 2D geometries, as the conditional phase shifter~\citep{Bertoni2000_PRL, Bertoni2002_JMO}, where the Coulomb interaction entangles electrons in different edge states.

\section*{ACKNOWLEDGEMENTS}
We thank Andrea Beggi and Chiara Galeotti for fruitful discussions. The work has been partially performed under the Project HPC-EUROPA3 (INFRAIA-2016-1-730897), with the support of the EC Research Innovation Action under the H2020 Programme; in particular, LB gratefully acknowledges the support of Departament d'Enginyeria Electr\`{o}nica (UAB) and the computer resources and technical support provided by Barcelona Supercomputing Center (project HPC17D8XLY). We also acknowledge CINECA for HPC computing resources and support under the ISCRA initiative (IsC57$\_$DYNAMET-HP10C1MI91). 
PB and AB thank Gruppo Nazionale per la Fisica Matematica (GNFM-INdAM). 
XO acknowledges funding from Fondo Europeo de Desarrollo Regional (FEDER), the 'Ministerio de Ciencia e Innovación' through the Spanish Projects TEC2015-67462-C2-1-R and RIT2018-097876-B-C21, the European Union's Horizon 2020 research and innovation program under grant agreement No Graphene Core2 785219 and under the Marie Skłodowska-Curie grant agreement No 765426 (TeraApps).

\section*{Appendix A. Split-Step Fourier Method for a two-particle system}
Together with a considerable increase in the memory cost, the evolution of two-particle wave function $\Psi(x_1,y_1,x_2,y_2;t)$ is characterized by an heavier computational load with respect to the single-particle version of the Split-Step Fourier method \citep{Beggi2015_JOPCM,Bellentani2018_PRB}. 

In presence of a perpendicular magnetic field, the linear momentum of the $i$-th particle is modified by including, in the Landau gauge, the magnetic vector potential $A_i=(0,Bx_i,0)$ with $i=1,2$, which separately couples with the $x_i$ and $k_{y_i}$ components of the single-electron hamiltonian. The two-particle Hamiltonian in presence of electron-electron interaction $\hat{V}_{12}$ then reads:
\begin{equation}
\hat{H}_{12}=\hat{V}_1+\hat{V}_2+\hat{V}_{12}+\hat{T}_{x_1,y_1}+\hat{T}_{x_2,y_2},
\end{equation} 
with $\hat{V}_{1},\hat{V}_{2}$ single-particle external potentials and 
\begin{equation}\label{eq:2PTB}
\hat{T}_{x_i,y_i}=\frac{\hat{p}^2_{x_i}}{2m^*}+\left( \frac{\hat{p}_{y_i}+eB\hat{x}_i^2}{2m^*}\right), \hspace{0.01\textwidth} i=1,2
\end{equation}
single-particle operator accounting for the dynamics of a free electron in a perpendicular magnetic field. The $\hat{x}$ and $\hat{y}$ components in Eq.~(\ref{eq:2PTB}) can be rearranged so that $\hat{T}_{x_1,y_1}+\hat{T}_{x_2,y_2}=\hat{T}_{x_1,x_2}+\hat{T}_{y_1,y_2}$, with
\begin{eqnarray}
\hat{T}_{x_1,x_2}&=&\frac{\hat{p}_{x_1}^2}{2m^*}+\frac{\hat{p}_{x_2}^2}{2m^*} \\
\hat{T}_{y_1,y_2}&=&\frac{\hat{p}_{y_1}^2}{2m^*}+\frac{2eB\hat{x}_{1}\hat{p}_{y_1}}{2m^*}+\frac{e^2B^2\hat{x}_1^2}{2m^*}+ \nonumber \\
&+&\frac{\hat{p}_{y_2}^2}{2m^*}+\frac{2eB\hat{x}_{2}\hat{p}_{y_2}}{2m^*}+\frac{e^2B^2\hat{x}_2^2}{2m^*}.
\end{eqnarray}
$\hat{T}_{x_1,x_2}$ is represented by a diagonal matrix in the 2D reciprocal space $[k_{x_1},k_{x_2}]$, regardless the space representation for the $\hat{y}$-coordinate, while $\hat{T}_{y_1,y_2}$ is diagonal in the 2D reciprocal space $[k_{y_1},k_{y_2}]$  and in the 2D real space $[x_1,x_2]$. On the other hand, the potential operators $\hat{V}_1$, $\hat{V}_2$ and $\hat{V}_{12}$ are diagonal on the 4D real space $[x_1,y_1,x_2,y_2]$. Note that $\hat{V}_{12}$ is a two-particle operator that couples the $\hat{x}$ and $\hat{y}$ coordinates: its exact representation is possible only in the 4D real space configuration $[x_1,y_1,x_2,y_2]$, where our two-particle wave function is defined.
By means of the approximate Trotter-Suzuky factorization \citep{Kramer2010_PS}, the evolution operator for a total evolution time $t=N\cdot\delta t$ can be separated into three terms:
\begin{equation}
[e^{-\frac{i}{\hbar}\delta t \hat{H}_{12}}]^N=[e^{-\frac{i}{\hbar}\delta t \cdot (\hat{V}_1+\hat{V}_2+\hat{V}_{12})}e^{-\frac{i}{\hbar}\delta t \cdot \hat{T}_{x_1,x_2}}e^{-\frac{i}{\hbar}\delta t \cdot \hat{T}_{y_1,y_2}}]^N.
\end{equation}
As in the standard Split-Step Fourier method\citep{Kramer2010_PS}, the diagonal character of the exponential operators in the 2D real/reciprocal spaces described above can be exploited in computing their effect on the quantum state by applying a 2D Fourier transforms $\mathcal{F}_{x_1,x_2}(\mathcal{F}_{y_1,y_2})$ and antitrasforms $\mathcal{F}^{-1}_{x_1,x_2}(\mathcal{F}^{-1}_{y_1,y_2})$ to the wave function, as in Eq.~(\ref{eq:evoloperat}).

\bibliography{HOM_BIBL}
\bibliographystyle{apsrev4-1}

\end{document}